\documentclass[twocolumn]{IEEEtran} 

\usepackage[pdftex]{graphicx}

\DeclareGraphicsExtensions{.pdf, .jpg, .tiff}

\begin{document}

\title{Three-dimensional structure of the laser wakefield accelerator in the blowout regime } 
\author{J. F. Vieira, S.F. Martins, F. Fi\'uza, R.A. Fonseca, L.O. Silva, C. Huang, W. Lu, M. Tzoufras, F. Tsung, W.B. Mori, J. Cooley, and T. Antonsen Jr. 
 \thanks{J. F. Vieira, S. F. Martins, F. Fi\'uza, R.A. Fonseca, L.O. Silva are with GoLP/CFP Instituto 
Superior T\'ecnico, 1049-001 Lisboa, Portugal 
(e-mail: jorge.vieira@ist.utl.pt)}
\thanks{C. Huang, W. Lu, M. Tzoufras, F. Tsung, W.B. Mori are
 with University of California, Los Angeles CA 90095, USA} 
 \thanks{J. Cooley, T. Antonsen Jr.  are
 with University of Maryland, College Park, MD 20742, USA}
 }
 
\markboth{March 2008}{Vieira {\it et al}: 3D Wakes}

\maketitle

\begin{abstract}
Three-dimensional Particle-in-Cell (PIC) simulations with the code QuickPIC are used to illustrate the typical accelerating structures associated with the interaction of an intense laser beam with an underdense plasma in the blowout regime. Our simulations are performed with an externally injected electron beam, positioned in the region of maximum accelerating gradients. As the laser propagates in the plasma, almost complete electron cavitation occurs, leading to the generation of accelerating fields in excess of 1~GeV/cm.
\end{abstract}




The recent developments on infrared ($\lambda_0\sim~\mu\textrm{m}$), ultra-intense ($>10^{19}~\textrm{W/cm}^2$), ultra-short ($< 30$~fs), and tightly focused ($\simeq 20~\mu\textrm{m}$) laser beams have opened the possibility to explore new scenarios on laser-plasma interactions, namely electron acceleration in the bubble~\cite{bib:pukhov} or blowout~\cite{bib:lu} regime of the laser wakefield accelerator (LWFA)~\cite{bib:tajima}. In many LWFA experiments \cite{bib:mangles,bib:geddes,bib:faure,bib:leemans} the laser intensity is sufficiently high for the radiation pressure of the laser to cause complete electron cavitation. Since ions are essentially stationary during a plasma wave period, a strong space charge is formed that pulls electrons back to the axis. The wake then assumes a spherical shape~\cite{bib:pukhov,bib:lu}, with linear accelerating and focusing forces. In addition to the non-linear plasma response, the laser self-modulations also play a key role in LWFA experiments. Thus, the overall laser-plasma dynamics is strongly coupled, and a complete theoretical understanding of typical LWFA experiments is beyond the capabilities of current pure theoretical models.

 
Simulations play an important role in this matter, by giving deeper insights on the physics of the LWFA, and being also valuable in the modeling and design of new experiments. In the pursuit of higher energy gains, lower plasma densities are required. Although the peak accelerating field is reduced for lower plasma densities with $E_{\mathrm{accel}}\propto\sqrt{n_e}$, where $n_e$ is the plasma density, the dephasing/pump-depletion lengths increase with $L_{\mathrm{accel}}\propto n_e^{-3/2}$. Therefore, the maximum energy gain increases with $W_{\mathrm{max}}\sim E_{\mathrm{accel}} L_{\mathrm{accel}}\propto n_e^{-1}$. The use of lower densities, however, constitutes a major difficulty to LWFA simulations using standard full PIC codes: it requires both finer grids to correctly resolve the laser wavelength, and more simulation time steps to model the increased interaction length. Therefore, the use of reduced PIC models to examine the laser-plasma interaction in the LWFA is becoming increasingly important. In this paper, we use the reduced PIC code QuickPIC~\cite{bib:chengkun} to illustrate the properties of the wake created by an intense laser pulse in the presence of an external electron beam. QuickPIC works under the quasi-static approximation (QSA)~\cite{bib:sprangle}. The QSA assumes that the typical time for the laser evolution is much larger than the typical time for the plasma response, precluding, however, the physics associated with self-injection. This assumption is very well verified in the LWFA. In QuickPIC, the plasma response is then calculated according a quasi-static field solver for each transverse plasma slice and the laser is advanced through the ponderomotive guiding center approximation. With this approximation larger time steps can be used for the laser evolution, thus leading to computational time savings of more than three orders of magnitude in comparison to standard full PIC codes.

\begin{figure*}[htb]
\begin{center}  
\includegraphics[scale=0.47]{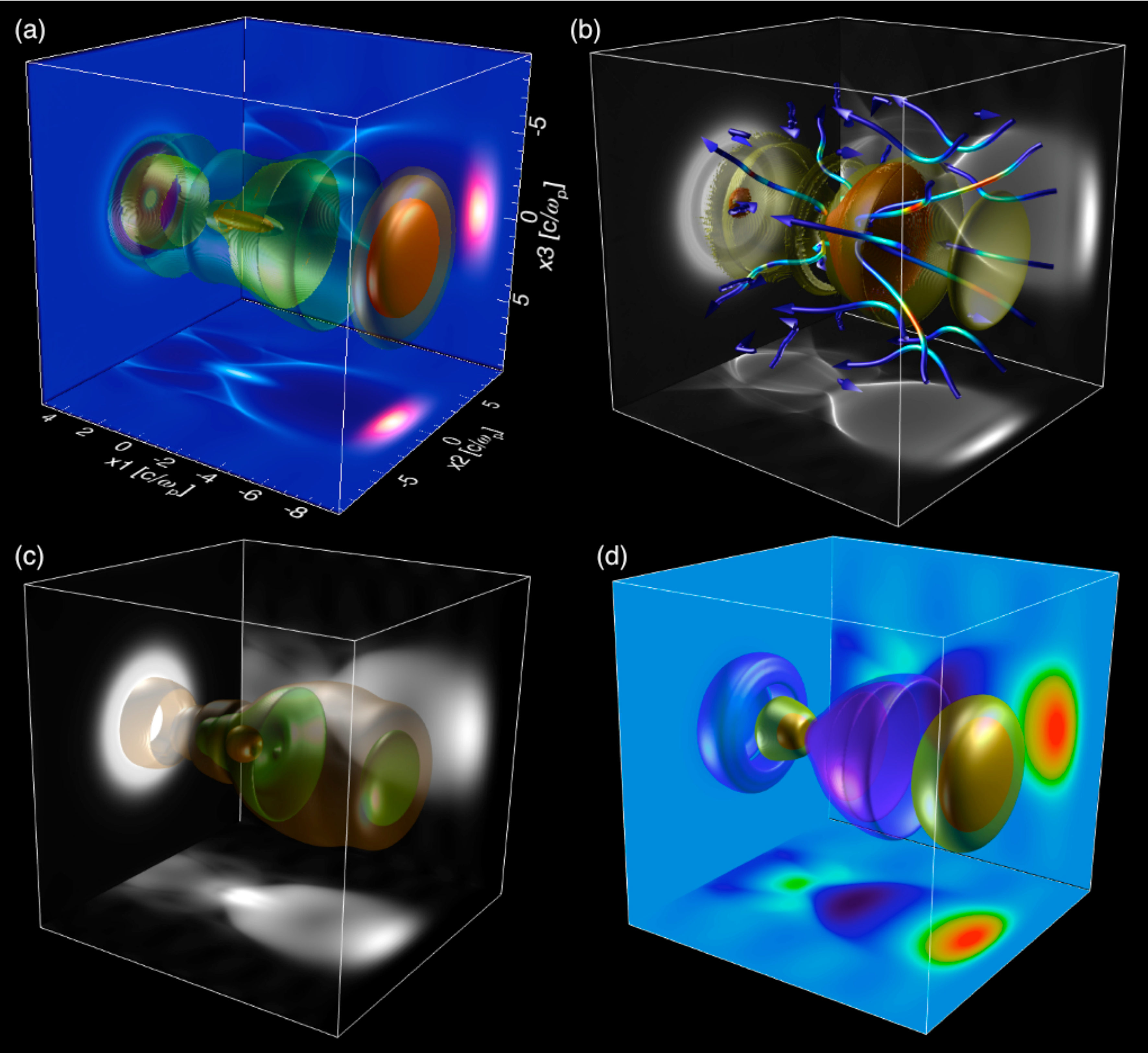}
\caption{\label{fig:fig1} (a) Isosurfaces of the plasma and externally injected beam charge density (green). The isosurfaces of the laser fields are shown in the front of the box (gold). The laser propagates in x1. The projection corresponds to the plasma and beam densities, and to the laser intensity. (b) Intensity isosurfaces and fieldlines (higher current values are in red and lower current values are in blue) of the plasma current in the blowout region. The projection corresponds to the intensity of the plasma current. (c) Isosurfaces of the focusing force, with the corresponding projections. (d) Isosurfaces of the longitudinal accelerating gradient. The projections also correspond to the accelerating wake field (higher field values are in red and lower field values are in blue)}
\end{center}
\end{figure*}

The three-dimensional structures associated with the laser propagation in the blowout regime are illustrated in Figure \ref{fig:fig1}. We use a laser pulse with a central wavelength $\lambda_0=0.8~\mu\textrm{m}$, focused to $20~\mu\mathrm{m}$, with a duration of 30~fs, and a peak normalized vector potential $a_0=4$. The laser pulse propagates in a plasma with density $n_e=10^{18}~\textrm{cm}^{-3}$. A bi-gaussian electron beam is externally injected, with a transverse width of $2~\mu\textrm{m}$, duration of 20~fs, charge of 1~nC, and traveling with a relativistic factor $\gamma=10^4$. The simulation box moves at the speed of light and it is resolved with $256\times256\times512$ cells for the transverse and longitudinal directions respectively, with 16 particles per cell. 

The plasma electrons are almost completely evacuated from the region where the laser is sitting (cf. Fig.~\ref{fig:fig1}-(a)). As they return to the axis (cf. Fig.~\ref{fig:fig1}-(b)), the plasma electrons feel the repulsive radial Coulomb force associated with the externally injected beam, leading to the flattening of the linear focusing force (cf. Fig.~\ref{fig:fig1}-(c)). As a consequence, the wake does not fully close at the end of the first plasma period. Similarly, in the longitudinal direction the accelerating gradient is also flattened by the presence of the electron beam (cf. Fig.~\ref{fig:fig1}-(d)). 

In conclusion, we have presented figures of three-dimensional waves in the blowout regime. These images clearly illustrate the fundamental structures associated with the LWFA, and allow a better understanding of the physics associated with these experiments.

\section*{Acknowledgments}
 This work was partially supported by Funda\c c\~ao para a Ci\^encia e Tecnologia  (Portugal) with grants SFRH/BD/22059/2005 and POCI/66823/2006. The simulations presented here were produced using the IST Cluster (IST/Portugal).

\nocite{*}
\bibliographystyle{IEEE}

%
\end{document}